# Emergence of near-boundary segregation zones in face-centered cubic multi-principal element alloys

Megan J. McCarthy [a,b], Hui Zheng [c], Diran Apelian [b], William J. Bowman [b], Horst Hahn [b,d], Jian Luo [c], Shyue Ping Ong [c], Xiaoqing Pan [b], Timothy J. Rupert [a,b,*]

[a] Materials and Manufacturing Technology Program, University of California, Irvine, CA, USA
[b] Department of Materials Science and Engineering, University of California, Irvine, CA, USA
[c] Department of NanoEngineering, University of California, San Diego, La Jolla, CA
[d] Institute of Nanotechnology, Karlsruhe Institute of Technology, 76344, Eggenstein-Leopoldshafen, Germany
* Email: trupert@uci.edu

Grain boundaries have been shown to dramatically influence the behavior of relatively simple materials such as monatomic metals and binary alloys. The increased chemical complexity associated with multi-principal element alloys is hypothesized to lead to new grain boundary phenomena. To explore the relationship between grain boundary structure and chemistry in these materials, hybrid molecular dynamics/Monte Carlo simulations of a faceted $\Sigma 11$ <110> tilt boundary, chosen to sample both high- and low-energy boundary configurations, are performed in face-centered cubic CrFeCoNiCu and CrFeCoNi equiatomic alloys. Unexpected enrichment of Fe is discovered in the face-centered cubic regions adjacent to the interface and found to be correlated with a structurally-distinct region of reduced atomic volume. Comparison with the boundary of the same type in monatomic Cu demonstrates that altered near-boundary regions exist in simpler systems as well, with the chemical complexity of the multi-principal element alloys highlighting its existence and importance.

# 1. Introduction

Research into multi-principal element alloys (MPEAs), also known as high-entropy or complex concentrated alloys, has greatly accelerated in the past decade due to the observation of highly desirable combinations of strength, ductility, and fracture toughness [1–3], even down to cryogenic temperatures [4,5]. Studies of the origins of these unique properties have shown them to be directly related to an MPEA's nanoscale structural and chemical characteristics. The rugged potential energy landscape created by having multiple elements in a random solid solution (RSS) has been demonstrated to affect dislocation structure and motion [3,6–10], vacancy formation and migration [11,12], and grain boundary structure [13,14]. MPEAs also should also exhibit complex segregation behavior which can alter material properties. For example, the formation and presence of domains of short-range order has been shown to mitigate the effects of radiation exposure [15,16] and increase the strength and hardness of MPEAs [10,17]. Second phase precipitation in MPEAs has also been documented as the mechanism behind extremely high radiation tolerance [18]. Such discoveries highlight potential pathways to novel material design strategies through careful tuning of MPEA composition, microstructure, and chemistry.

In traditional alloys (i.e., those with a single principal element and smaller amounts of alloying additions), interfacial structure and chemistry are intimately connected. The character of grain boundaries has been shown to strongly influence the amount of segregant and the precise sites which may be occupied [19–21]. Similarly, segregating elements may themselves change the structure of interfaces [21–27], as has been observed in boundaries in the Cu-Zr [21], Cu-Ag [23], Pt-Au [25], and Ni-Bi [24] systems. There are many open questions about how such interfacial phenomena are expressed in MPEAs. The increased chemical complexity of MPEAs means that multiple segregating species can compete for sites and potentially heavily influence local ordering. Though this research topic is still in its nascent stage, there have been important advances. Atom probe tomography studies by Li and coworkers [28] on the Cantor alloy (equiatomic CrMnFeCoNi and its quaternary and ternary variants) have confirmed the influence of grain boundary macroscopic orientation (i.e., the grain boundary character) and microscopic structure on segregation and grain boundary energy, as has been well-established in traditional alloys [22,29]. The resulting segregation may also act as a potential pathway to phase transformations at boundaries [30], indicating a potential means of augmenting or even transforming grain boundary structure in MPEAs.

Microstructural engineering for MPEAs will require a deeper exploration of the specific mechanisms behind nanoscale segregation behavior at defects. Atomistic simulations are uniquely suited to this task and have already shown promise as important tools in MPEA research on the Cantor alloy. For example, Chatain and Wynblatt [31,32] have demonstrated how intricate interactions between elements result in layered compositions at grain boundaries and surfaces in this alloy. Their analyses were also used to connect atomistic simulation results to larger-scale analytical models. Hu and Luo [33], using atomistic simulation combined with machine learning, have revealed complex and unique co-segregation relationships between sets of elements and the degree of disorder in grain boundaries. Such techniques not only quantify important fundamental relationships between segregating elements, but also provide a means of exploring the enormous compositional space of potential MPEAs, one of the most important challenges in this area [1]. To build on this prior work, fundamental studies of other face-centered cubic (FCC) MPEAs beyond the Cantor alloy will be important for clarifying the role of factors such as atomic size and elemental energetic properties in segregation between different MPEA configurations.

In this work, atomistic simulations are used to explore segregation and its effect on the structure of a faceted $\Sigma11$ boundary in two different FCC MPEAs: a quinary alloy (CrFeCoNiCu) and a quaternary alloy (CrFeCoNi). We find that even strong segregation does not alter the fundamental structure of the boundary with respect to the RSS or pure configurations, indicating that the FCC crystal structure is more important to faceted boundary morphology than local composition. In the quinary alloy, Cu dominates segregation and in almost every case leads to the depletion of all other elements, resembling the simpler behavior of a binary alloy system. The quaternary variant of the MPEA has more subtle trends in its segregation, with Cr and Co both exhibiting strong segregation tendencies. In both materials, Ni and Fe are the most depleted within the boundary plane itself. There are unexpected enrichments of Fe in the FCC-ordered regions directly adjacent to facet planes and junctions in both alloys, marking the discovery of a near-boundary segregation region. These enrichments are found to be correlated to a zone of reduced atomic volume next to the boundary which, while discovered in the chemically-complex MPEAs here, we find is also present in pure Cu grain boundary structures. Relationships between per-element atomic volumes, changes in composition, and grain boundary segregation trends are uncovered and explored.

## 2. Methods

All simulations in this study were run using the Large-scale Atomic/Molecular Massively Parallel Simulator (LAMMPS) software package [34]. Atomic volumes were calculated in LAMMPS using the library Voro++ [35]. Data analysis and visualizations were performed using the OVITO software [36]. Atomic interactions for all samples (elemental metal, quinary MPEA, and quaternary MPEA) were modeled with the embedded-atom method (EAM) potential created by Farkas and Caro [37] for the CrFeCoNiCu system. Using the melt interface tracking method outlined by Wang et al. [38], the melting temperatures for the quinary and quaternary alloys were determined to be approximately 2090 K and 2040 K respectively, with an estimated error of ±20 K for each. An additional grain boundary in pure Cu (shown in Section 3.3) was generated using the embedded-atom method potential by Mishin et al. [39].

The selection of the Farkas and Caro potential was motivated by a desire to explore grain boundary segregation in FCC MPEAs outside of the Cantor alloy, while also taking advantage of the computational efficiency of an EAM potential. Some properties of this potential require a brief explanation. In order to form a stable, single-phase FCC random solid solution through all temperatures, the potential fitting procedure minimized the total heat of mixing of the random solid solution, leading to pairwise enthalpies of mixing that are also close to zero at 0 K. As part of this process, the pure element properties at T = 0 K, such as average pure atomic volumes (shown in Table 1) were fit based on FCC crystal structures. We note here that, because atomic volume will be an important parameter used in the discussion of our results, we use average pure atomic volume as a proxy for an element's atomic radius or lattice parameter. While FCC is the natural crystal structure for Cu and Ni, at low temperatures, Co typically has a hexagonal close packed (HCP) crystal structure, while both Cr and Fe are body-centered cubic (BCC). Table 2 includes the reported enthalpies of mixing, along with additional calculations for each binary alloy pair. Simulation cells with 50% of each element per pair were run at T = 1000 K and a constant pressure of zero, using a hybrid molecular dynamics (MD)/Monte Carlo (MC) simulation (additional details below). The enthalpies of mixing were extracted from the cells after 0.5 ns of simulation time, which for this potential was enough to ensure the creation of a FCC binary solid solution. These values are also comparatively low, with a maximum magnitude of +1.359 kJ/mol for the pair Ni-Co. Low magnitudes of enthalpy are not unexpected, as the replication of the FCC bulk phase of MPEAs was of primary importance to the authors of the potential and is also a main feature of interest in the current study.

| | Cr (FCC) | Fe (FCC) | Co (FCC) | Ni | Cu |
|---|---|---|---|---|---|
| $\bar{V}_{atom}$ [Å³], pure bulk | 11.00 (4) | 11.28 (2) | 11.18 (3) | 10.90 (5) | 11.86 (1) |
| $\bar{V}_{atom}$ [Å³], quin. bulk | 11.78 (2) | 11.66 (5) | 11.67 (4) | 11.69 (3) | 11.95 (1) |
| $\bar{V}_{atom}$ [Å³], quat. bulk | 11.67 (1) | 11.55 (4) | 11.56 (3) | 11.59 (2) | - |
| Relaxed SFE [mJ/m²] | 24 | 43 | 15 | 125 | 43 |
| $E_{vacancy}$ [eV/atom] | 1.41 | 1.61 | 1.36 | 1.61 | 1.19 |
| $E_{110}$ [J/m²] | 2.02 | 2.00 | 2.08 | 2.05 | 1.31 |
| $E_{100}$ [J/m²] | 1.88 | 1.86 | 1.90 | 1.88 | 1.25 |
| $E_{111}$ [J/m²] | 1.65 | 1.70 | 1.60 | 1.75 | 1.24 |
| $E_{GB}$, Σ11 [mJ/m²] | 522 (3) | 541 (4) | 521 (2) | 610 (5) | 309 (1) |

**Table 1.** Selected properties of the interatomic potential. For each element listed, $\bar{V}_{atom}$ is the average atomic volume of each element measured in the bulk of a pure cell and the quinary and quaternary alloys, the relaxed SFE is the relaxed stacking fault energy, $E_{vacancy}$ is the vacancy formation energy, the $E_{001}$, $E_{110}$, and $E_{111}$ are the surface energies for the (001), (110), and (111) planes respectively, $E_{GB}$, Σ11 is the grain boundary energy for the symmetric Σ11 grain boundary (i.e., the low energy facet) alone. The values in parentheses in the $E_{GB}$, Σ11 row indicate the ranking of lowest (1) to highest (5) grain boundary energies. The values for average atomic volume and $E_{GB}$, Σ11 were calculated by the authors. All other values were taken from data published by Farkas and Caro [37].

| Material | Pair | $\Delta H_{A-B}$, RSS [37], T = 0 K [kJ/mol] | $\Delta H_{A-B}$, MD/MC, T = 1000 K [kJ/mol] |
|---|---|---|---|
| **Quinary only** | Cu-Cr | 0.521 | 0.337 |
| | Cu-Ni | 0.703 | 0.339 |
| | Cu-Co | 0.428 | 0.073 |
| | Cu-Fe | 0.374 | -1.008 |
| **Quinary and quaternary** | Co-Ni | 0.135 | 1.359 |
| | Fe-Ni | 0.176 | 0.470 |
| | Cr-Fe | 0.393 | 0.375 |
| | Co-Fe | 0.109 | 0.182 |
| | Co-Cr | 0.144 | 0.133 |
| | Ni-Cr | 0.095 | 0.060 |

**Table 2.** Enthalpies of mixing $\Delta H_{A-B}$ for binary alloy pair A-B, each of which is present in the quinary and quaternary alloys. The first column contains enthalpy data for the random solid solution (RSS) and the second column data from calculations of mixing enthalpies for each equimolar binary alloy pair.

A faceted $\Sigma 11$ <110> tilt boundary was chosen for this study in order to efficiently sample different grain boundary structures in one dataset. Grain boundary faceting is a process in which an initially-flat interface spontaneously dissociates into a periodic array of two different segments, one with low energy and one with higher energy. The low and high energy facets are oriented along different planes, meaning that the sites where they join introduce geometrically-necessary defects in the boundary structure. These defects, called *facet junctions*, are interesting in and of themselves as sites with important geometric properties regulating facet structure [40,41], as well as sites of preferred segregation [42]. The faceted $\Sigma 11$ boundary is of particular interest for having facets with interesting structural features [43–45], special defects [44,46], and dynamic behavior [46,47], as well as a well-developed body of literature from chemically simpler materials [11,43–45,48]. Example snapshots of the $\Sigma 11$ faceted boundary in pure Cu is shown in Fig. 1(a) and (b). All boundary snapshots shown are viewed down the Z-direction, parallel to the tilt axis of <110>. To reduce thermal noise in higher-temperature conditions, a minimization step was performed on each sample prior to visualization. Figure 1(a) presents a map of the potential energy (eV/atom) of this boundary, highlighting the positions of the low and high energy facets. The orientations of each facet's terminal planes are also designated on this figure. Dark blue coloring indicates atoms with potential energies near the average energy of cohesion $E_{coh}$ for Cu in this potential (-3.54

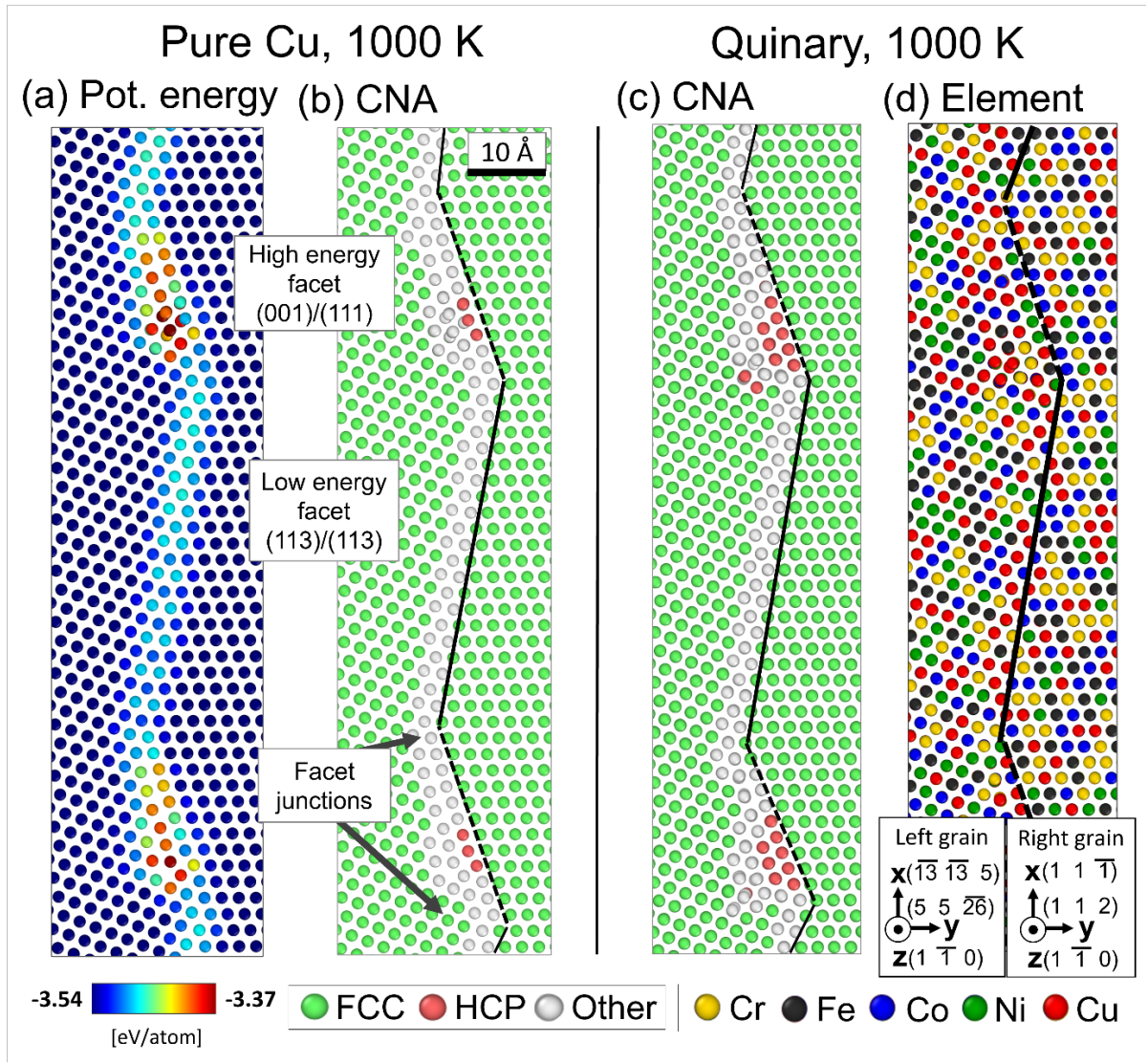

**FIG 1.** Examples of the faceted Σ11 <110> tilt boundaries. (a, b) The boundary in pure Cu, with atoms colored by potential energy (eV/atom) in (a) and colored by Common Neighbor Analysis (CNA) in (b). The high energy facets have groupings of yellow to red colored atom and are denoted with a dashed line in (b), while the low energy facets have light blue to dark blue atoms in (a) and are denoted with a solid line in (b). (c) The same type of boundary in the quinary MPEA (CrFeCoNiCu) at 1000 K after 1 ns of MD/MC simulation. (d) A view of the elemental composition of the boundary in (c), which includes the boundary plane indices.

eV/atom, Table 1) and corresponds to bulk atoms and atoms in other low energy configurations.

The high energy facets are quickly identified by regions of orange and red colored atoms (averaging around -3.37 eV/atom), while the atoms of the low energy facets remain in the light blue to medium blue color range (with the highest values at approximately -3.48 eV/atom) of the potential energy spectrum.

To show how the regions of low and high energy in Fig. 1(a) correspond to boundary structure, Fig. 1(b) shows the same boundary now colored using OVITO's Common Neighbor Analysis (CNA) algorithm [49]. As indicated by the accompanying legend at the bottom of Fig. 1, green coloring indicates FCC coordination, red coloring indicates HCP coordination, and white coloring indicates "other" coordination, which in this study almost exclusively correlates with grain boundary atoms. The solid and dashed black lines on Fig. 1(b) are parallel to the different plane orientations of the low and high energy facets, respectively. The low energy facet is oriented along the symmetric boundary plane of the $\Sigma 11$ coincident site lattice, with terminal planes of (113)/(113). The high energy facet is oriented along two different planes for each grain, namely the (001)/(111) planes of the left and right grains, respectively. This type of interface is called an incommensurate boundary plane, which has a boundary plane spacing that is an irrational number ($\sqrt{3}/1$ in this case). A label with arrows indicates two examples of facet junctions. By studying combinations of high and low energy boundary regions, we can ensure that any observed segregation trends are general, rather than associated with a specific or special boundary type.

The specific structure of grain boundaries created in atomistic simulations can be sensitive to the method of creation. To ensure uniformity and reproducibility, the bicrystals shown in Fig. 1 were generated using an iterative method developed by Tschopp et al. [50], used to locate minimum grain boundary energy configurations. This method involves first choosing the desired grain orientation and final length of the cell normal to the grain boundary ($L_y$ in this work). Using this information, the algorithm calculates a minimum system size in the other two dimensions to allow fully periodic boundary conditions, creating two grain boundaries per cell. One of the grains is then systematically shifted relative to the other and a conjugate gradient minimization is performed, generating a large set of boundary configurations with varying energies. These configurations spontaneously facet as a part of the simulated MD relaxation process, arising from the choice of boundary plane orientation (plane indices indicated in Fig. 1(d)). More details of the geometric and energetic conditions that lead to faceting in $\Sigma 11$ boundaries can be found in Brown and Mishin [44]. As the generated cells are deliberately minimized in the directions containing the grain boundary plane ($L_y$ and $L_z$ in this case), the chosen cell is replicated in those dimensions to reach the desired size ($L_x, L_y, L_z \sim 270$ Å, 200 Å, 36 Å). The replication procedure results in a structure with 162,288 atoms for this study.

The algorithm from Ref. [50] was not designed for multi-component systems, which would technically also require chemical relaxations (i.e., the sampling of thousands of different elemental configurations) to find the absolute minimum energy structure, so the $\Sigma 11$ bicrystals were created using a pure element. Because of faceted boundaries' unique morphologies, choice of the pure cell's element required some consideration. Cu and Ni are natively FCC, making them good initial candidates for cell generation. A previous study on faceted $\Sigma 11$ boundaries using six different FCC potentials showed the influence of stacking fault energy on faceted boundary morphology and properties [46]. Between the two elemental choices, pure Cu (from the same interatomic potential) has the closest relaxed stacking fault energy to the reported average for CrFeCoNiCu (50 mJ/m$^2$) and CrFeCoNi (51 mJ/m$^2$). Pure Cu was therefore chosen to create the initial computational cells, after which the atomic types were randomized to create an equiatomic random solid solution (RSS) as the starting point for further simulation.

To create the fully segregated quinary and quaternary alloy grain boundaries, a hybrid MD/ MC algorithm was applied to the RSS bicrystals. A Nosé-Hoover thermostat and barostat was used to control temperature (varies) and pressure (zero) with a timestep of 1 fs. To equilibrate boundary structure with an annealing step, kinetic energy equivalent to approximately half the target temperature was added to the atoms in the cell and then the temperature was ramped over 20 ps up to the target value. The structure was then annealed for a further 30 ps. After this initial equilibration, the hybrid MD/MC method was applied to the structure to allow segregation to and relaxation of the grain boundaries. MC steps were run using the variance-constrained semi-grand-canonical (VC-SGC) ensemble from Sadigh et al. [51]. This ensemble, an extension of the traditional semi-grand-canonical (SGC) ensemble, was developed in order to enable large-scale, time-efficient simulation of alloys in which multiphase precipitation is known to occur or is hypothesized. In the context of this work, the VC-SGC MC method was chosen in order to allow for the possibility of such phase transitions. For example, experimental studies have indicated a tendency for the quinary MPEA to separate into two FCC phases, one Cu-rich and the other Cu-poor [52]. Traditional SCG algorithms are only able to stabilize a single phase and would not be able to replicate this two-phase material, making the use of VC-SGC necessary. In addition, grain boundary segregation can result in phase-like changes within grain boundaries [27], known as complexion transitions, a possibility which must also be tested using VC-SGC. Prior work, such as that of Frolov et al. [22,23] have shown that, even in binary alloys, segregation can induce such transitions within grain boundaries. The presence of two facets (i.e., two boundary planes with very different character) in a material with complex chemistry could, in theory, lead to different segregation trends and correspondingly different capacities for similar interfacial structural transitions.

In VC-SGC MC simulation steps, MC transmutation is carried out as it would be in a SGC simulation, but the change in composition is tightly controlled (i.e., the variance in composition is constrained). The difference in pair-wise chemical potential energies between elements A and B, $\Delta\mu_{AB}$, dictate the MC acceptance/rejection probability. Appropriate values for $\Delta\mu_{AB}$ were found using an iterative algorithm optimizing the acceptance rate of atom swaps and potential energy reduction per temperature. The variance constraint parameter, $\kappa$, was set to $10^3$, which keeps the standard deviation of concentration values less than 0.05% away from the target concentration. For every 20 MD steps, one MC step was performed, which randomly selected 25% of the atoms in the cell to attempt to swap. Each simulation was run for 1 ns, leading to a total of 50,000 MC steps. At least three runs were completed for each combination of material and temperature, initiated using three unique but equivalent numerical seeds for kinetic energy randomization. In addition, sampling from every 100 ps of MD simulation time (alternatively, every 5,000 MC steps) starting from 400 ps of each MD/MC run gives a total of 21 samples per combination. Boundary snapshots of the segregated structures are taken from the final timestep (that is, the segregated state at 1 ns), unless otherwise indicated.

Figure 1(c) and (d) show an example of the CNA structure and elemental distribution, respectively, of a quinary MPEA boundary after 1 ns of MD/MC simulation at 1000 K. Since grain boundary segregation in traditional alloys has been demonstrated to induce fundamental transformations in boundary properties [25], it is important to establish whether any such transition has occured after segregation in boundaries in MPEAs. Confirmation is especially important in faceted boundaries, which have been shown to be structurally sensitive to segregation. In a hybrid experimental/computational study, Priedeman and Thompson [25] showed how Au segregation to faceted Pt boundaries can alter facet periodicity, a fundamental structural property of faceted boundaries in pure materials.  As can be seen by comparing the pure Cu boundary in Fig. 1(a) and (b) to the fully segregated quinary shown in Fig. 1(c) and (d), there is no evidence of structural transformations in the $\Sigma11$ boundary studied here.  The same can be said for quaternary alloy (not shown), which strongly resembles the example shown in Fig. 1(c).

## 3. Results and Discussion

### 3.1 General grain boundary segregation trends

The local concentration of each element, $C$ (at.%), versus position near the grain boundaries in the quinary alloy is shown in Fig. 2(a), with this figure including all atoms in the grain boundary regions (i.e., including both facet types).  The value $C$ is calculated by binning the atoms in the bicrystal along the Y-direction (bin size of approximately 2 Å), calculating the percent of atoms of a given element in each bin.  The center of the plot is located at the mean boundary position, with data captured and averaged for both the upper and lower boundaries in the bicrystal.  A small schematic of the orientation used in this analysis is included as an inset. The colored lines indicate the average fraction for each element, while the shaded bands around them show the standard deviation.  At this temperature, Cu is the dominant segregating element in the quinary system, with a maximum average value of 29.4 at.% as compared to the initial concentration of 20 at.% for all components.  Cr, Ni, and Fe are depleted from the boundary, with the latter two being depleted most strongly.  Of the four remaining elements, only Co remains near the initial concentration within the boundary.  The strong Cu segregation has led to mild depletion of Cu in the bulk (approximately 19.4 at.% on average), a phenomenon predicted by Wynblatt and Chatain [31] for alloys with strongly-segregating elements in polycrystals with grain sizes beneath 100 nm. Figure 2(b) shows the grain boundary composition, taken by measuring the composition of all non-FCC atoms, as a function of temperature from 200 K to 1800 K, which correspond to homologous temperatures of approximately 0.1 to 0.9. Increasing temperature decreases the amount of Cu segregation in an Arrhenius-like manner, which is consistent with what would be expected from a Langmuir-McLean model of grain boundary adsorption [53,54]. We note here that the intermediate temperature of 1000 K (black arrow) is chosen for Fig. 2(a) and will be used for all composition and structural analysis moving forward. At this temperature, facets form with a minimum of defects, while still allowing for significant segregation in both alloys.

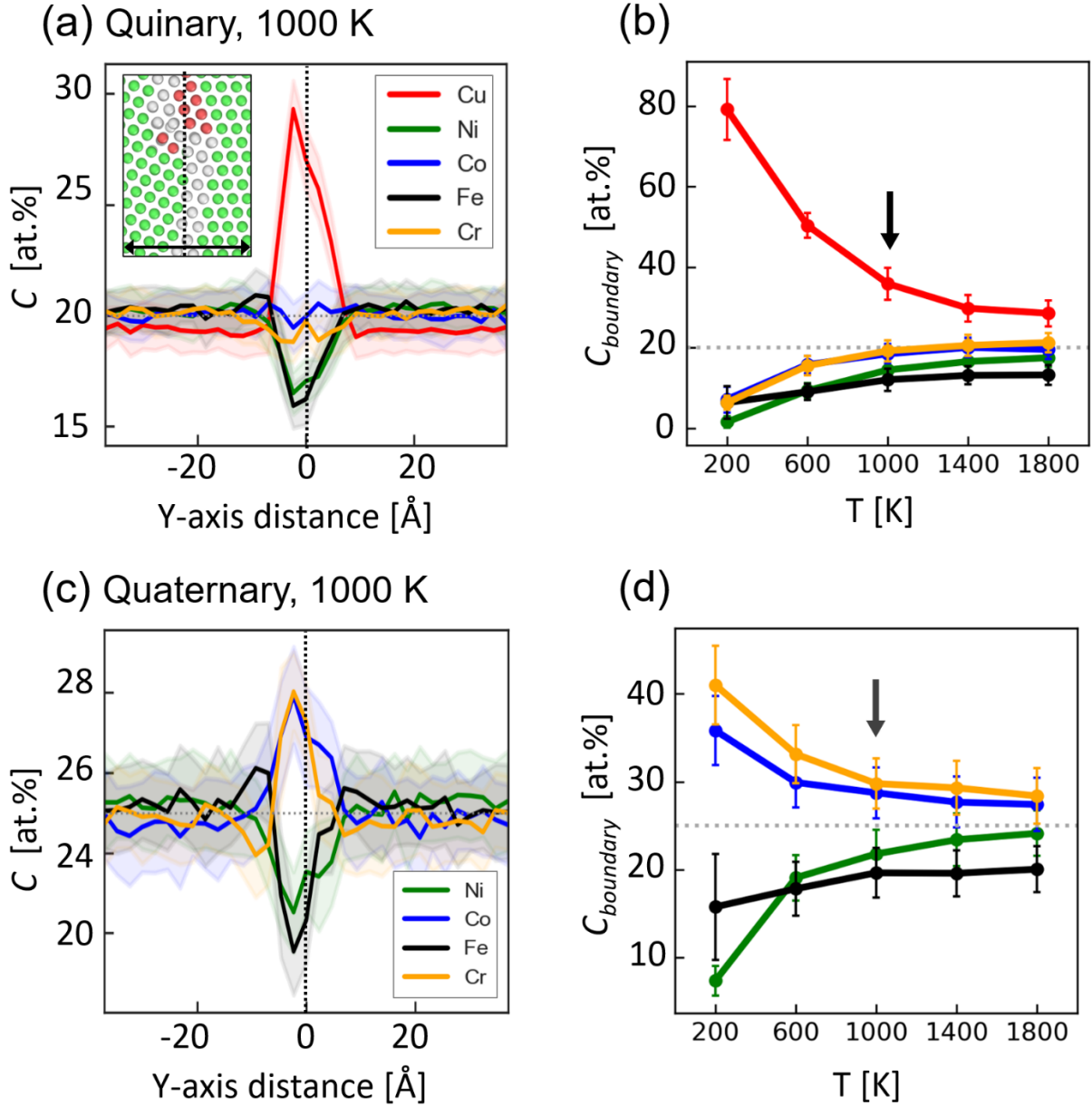

**FIG 2.** (a) Local concentration as a function of Y-position data reflecting the segregated state of the grain boundary in the quinary alloy at 1000 K. The inset provides the orientation for the direction of measurement. (b) Grain boundary concentration (i.e., non-FCC atoms) as a function of temperature for the quinary alloy, demonstrating Arrhenius-like decreases in segregation of Cu with increasing temperature. (c) Local concentration as a function of Y-position for the quaternary alloy (without Cu). (d) Grain boundary concentration as a function of temperature for the quaternary alloy, showing both Cr and Co experience decreases in concentration with increasing temperature. The black arrows in (c) and (d) indicate the chosen temperature for detailed structural studies. All error bars/shaded bands in this and following figures show the standard deviation of each measurement.

Important in the study of any MPEA is an understanding of its chemical short-range order (CSRO), which can have a large impact on material properties [17,55–57]. One commonly-used tool for describing CSRO is a variant on the Warren-Cowley equation [58], which counts the relative frequency, $\alpha_n$, of atom pairs using the $n$th nearest-neighbor shells of a material:

$$\alpha_n \ = \ 1 - \frac{P_n^{ij}}{c_j} \tag{1}$$

On the left-hand side, the subscript $n$ refers to the neighbor shell of interest (here, always the 1$^{st}$). On the right-hand side, the term $P_n^{ij}$ describes the probability of finding an atom of species $j$ in the $n$th shell of an atom of species $i,$ and the term $c_j$ denotes the global concentration of atoms of type $j$. A value of zero corresponds to a material resembling an ideal RSS. Negative values indicate that pairs are more likely to be found in each other's $n$-th nearest-neighbor shell (i.e., clustering behavior), while positive values reflect a lower frequency of pairs in the same $n$-th shell (i.e., anti-clustering). Figure 3 shows the $\alpha_1$ values for the first nearest-neighbor shells for the bulk material (gray bars)and for the Σ11 boundary atoms (colored bars) for each alloy. The bulk values indicate that almost all atoms outside the boundaries have $\alpha_1$ values very close to zero at 1000 K, indicating that bulk atoms do approximate an RSS in this system. In contrast, the CSRO values for the grain boundary atoms are larger in magnitude than those of the bulk, with the mean $\alpha_1$ value of the quinary Cu-Cu pair -0.51 in Fig. 3(a) and all other pairs falling within the range of -0.2 to +0.3.

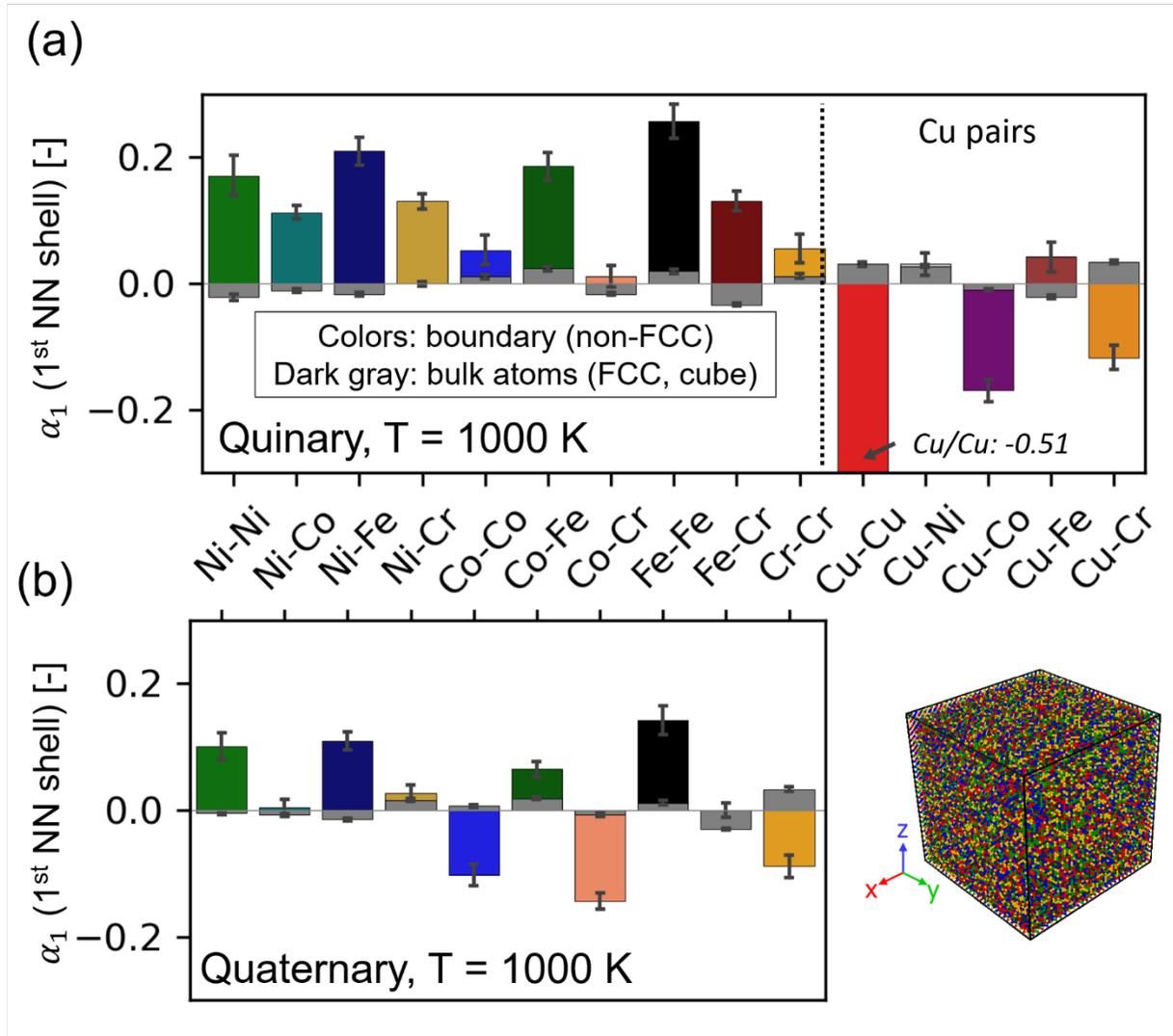

**FIG 3.** Chemical short-range order (CSRO) analysis for the quinary (top, a) and quaternary (bottom, b) alloys using the first nearest-neighbor shell $\alpha_1$ values for the Warren-Cowley parameter (Equation 1). Analyses were conducted both on a bulk cube (gray bars) and the grain boundary/non-FCC atoms (colored bars).

To better understand the driving forces behind elemental segregation, the theoretical model developed by Wynblatt and Ku [19,59] will be applied. Their solute segregation model was developed to approximate the behavior of binary alloys. As such, it is used in this work to provide a useful framework for the discussion of per-element behavior, where the element in focus is the solute and the surrounding RSS of the other elements is the solvent/matrix. Treating the random solid solution as a single material is a variation of the averaged-atom model applied to MPEAs in other contexts [13,14,31,60,61], a technique used to reduce complexity and gain insight into material trends. In this way, the properties of the bulk in MPEA samples which remain randomly distributed (as they do in these MPEAs) can be used as a reference point to understand changes at interfaces due to chemical relaxation (i.e., from MD/MC simulation). At the time of this writing, some promising models for multi-element segregation in MPEAs are being developed, such as that by Zhou et al. [62]. However, extensive development of their model would be necessary for use in this system due to the different interfacial bonding characteristics and the presence of irregularities in the high energy facet. In the Wynblatt-Ku model, the segregation enthalpy has (1) a boundary energy term, (2) an alloy interaction term, and (3) a strain energy term:

$$\Delta H_{seg} = -\frac{24\pi B \mu r_I r_M (r_I - r_M)^2}{3B r_I + 4\mu r_M} + (\sigma_I - \sigma_M) A^\phi - \frac{2\Delta H_{mix}}{Z X_I X_M} \left[ Z_L \left( X_I^\phi - X_M \right) + Z_P \left( X_I - \frac{1}{2} \right) \right] \qquad (2)$$

The subscripts $I$ and $M$ denote the solute and solvent contributions, respectively, and terms without those subscripts refer to properties of the solvent element. The first term describes the strain energy contribution to the segregation enthalpy, defined in terms of the solvent bulk modulus, $B$, the solvent shear modulus, $\mu$, and the differences in atomic radii, $r$, between the pair of elements. The second term in Equation 2 describes the driving force for segregation induced by the difference in grain boundary energy, $\sigma$, for the solute and solvent, multiplied by the interfacial area per volume, $A$. The remaining third term quantifies the influence of alloy interactions, or in other words, the contribution due to the chemical interactions of elemental pairs. $\Delta H_{mix}$ is the enthalpy of mixing for solute-solvent (I-M) pairs, $X_i$ describes the concentration of a component at the interface, $X_I^\phi$ describes the concentration of the solute within the boundary, $Z$ is the total coordination number per atom, and $Z_i$ represents the coordination number in different regions (L within its own plane and P across-plane). Other elemental interactions not described by Equation 2, such as site competition, could also be active and will be explored in upcoming sections.

The third term in Equation 2 is expected to have a small effect on segregation in these alloys, as the interatomic potential is purposefully designed to keep values of $\Delta H_{mix}$ very low. Again, while this choice is reasonable since the potential is intended to model a single FCC phase, it is possible that MPEAs with a significant tendency for the formation of short-range order would have a larger contribution from the third term, as this ordering is itself a sign that some combination of mixing enthalpy is strongly negative. The results from the bulk analysis of CSRO in Figs. 3(a) and (b) confirm that, outside of the grain boundaries, there is no significant ordering tendency. This means that the larger $\alpha_I$ values seen within the boundary can be mostly attributed to segregation induced by the first two terms of Equation 2, rather than chemical effects. A clear example of this correlation can be seen in the quinary in Fig. 3(a), where Cu, the most dominant segregating species by far, also has the most negative value. The only other pairs that are clearly negative are Cu-Co and Cu-Cr, the second- and third-most common element retained at the boundary, while all others are positive (indicating depletion).

With respect to the first two terms of Equation 2 above, in the quinary alloy Cu is by far the most favored element for segregation. Starting with the strain energy term, Cu atoms have by far the largest average pure atomic volume (again, serving as a representation of atomic radius or lattice parameter) for the FCC structure of any element in this potential (Table 1) and will therefore have the largest negative strain energy term for segregation enthalpy. A more thorough discussion of the importance of free volume at the grain boundaries, which is the root cause of the strain energy term, will follow later. The basic defect energies associated with Cu atoms also promote its segregation. Though surfaces and vacancies are different defects than grain boundaries, they are all expressions of the energy penalties resulting from dangling bonds. Because Cu has by far the lowest average values for $E_{vacancy}$, $E_{100}$, $E_{110}$, and $E_{111}$ (Table 1), defects with aggregations of Cu atoms will also tend to have lower energy penalties as well. Calculations of the $\Sigma 11$ boundary energy (i.e., the low energy facet by itself) for each material, shown in the last row of Table 1, confirm this trend, as Cu has the lowest energy at 309 mJ/m$^2$. Thus, Cu as the dominant boundary segregant is consistent with the second segregation enthalpy term as well.

The local composition versus position data and the temperature trend data for the quaternary alloy are shown in Figs. 2(c) and (d), where we remind the reader that the equiatomic bulk composition is now 25 at.%. Through all temperatures, Cr and Co both become roughly equally enriched, while Ni and Fe maintain their roles as the most heavily depleted elements as in the quinary alloy. With respect to the first term of Equation 2, there is no immediately clear explanation for strain energy-induced Cr and Co co-segregation. This is likely tied to the fact that the relative differences in atomic average pure atomic volume between the four elements in the quaternary are relatively small (a maximum difference of 0.08 Å$^3$ and a minimum of 0.01 Å$^3$). Even these differences would tend to favor Fe and Ni segregation over Cr and Co, which have the second- and fourth-largest average atomic volumes in the quaternary alloy. The second term of Equation 2 (boundary energy) offers more clarity. The $\Sigma 11$ grain boundary energies (last row of Table 1) indicate that Cr and Co have nearly identical values and are both lower than Fe and Ni. Additionally, both Cr and Co have lower energies for vacancy formation and (111) plane energies (both relevant for the more defect-dense high energy facet) than those of Fe and Ni by a significant margin. These energy trends suggest that segregation would be favored for Cr and Co over Ni and Fe, which is exactly what is observed in the quaternary. Additionally, Cr segregation is consistent with the first-principles calculations on this quaternary alloy conducted by Middleburgh et al. [63], who found that Cr is not thermodynamically stable in solution and would have a tendency to segregate. Thus, the energetic trends alone may provide the primary driving force for the co-dominance of Cr and Co segregation, with minor contributions from the first term.

While strong depletion of Fe and Ni appears to be characteristic for both MPEAs, the composition data of the quaternary MPEA specifically highlights that certain element pairs tend to have coupled segregation relationships. For example, in the regions ±10 Å from the mean boundary position, the Fe enrichment curve is closely mirrored by depletion in Cr with a similar magnitude. Although more subtle, Co and Ni show similar inverse changes. Such anti-symmetric couplings strongly suggest that the pairs Fe/Cr and Co/Ni are engaging in site competition with one another. Importantly, these coupled regions extend relatively far outside of the boundary plane, suggesting that segregation trends beyond simply the grain boundary itself may need to be considered in MPEAs. This topic will be explored further in an upcoming section.

### 3.2 Spatial variations in boundary composition for different facets

In the previous section, some general segregation trends of the quaternary and quinary alloys were established, using the Wynblatt-Ku model to frame the results. However, it is not yet clear how the low and high energy facets (i.e., more symmetric and asymmetric boundary planes) contribute to these trends. To understand the behavior of each facet type (noting that there are multiple facet lengths per gain boundary in the bicrystal), a plane-by-plane chemical analysis was conducted on each, with the results shown in Fig. 4. Starting with the central plane marked as Plane 0, each neighboring plane is numbered with a positive or negative integer indicating its relative position. Instructive examples are denoted for Planes -5 and 5 as dashed lines in Figs. 4(a) and (b). Because the high energy facets are asymmetric, their planes are not perfectly perpendicular with the grain boundary plane normal direction Y'', and they also contain some irregular defects on the left side. To account for these features, each plane was measured not directly parallel to the coordinate system, but at a slight angle (~3°) for the high energy facet. It is also important to note that the plane numbers between the low and high energy facets correspond to different spatial distances. For example, the region between Planes -5 and 5 in the low energy facet spans ~11 Å, while the same plane number range in the high energy facet spans ~17 Å. The resulting analysis of elemental composition versus plane number for the quinary and the quaternary alloys are shown in Figs. 4(c-f). The low energy facets are shown in the left column while the high energy facets are shown in the right column. While the overall trends from the boundary composition data from the previous analysis in Fig. 2 are the same, additional nuance can be observed. For example, one clear difference is the highly symmetric and ordered patterns of enrichment and depletion corresponding to each plane in the low energy facets. Such patterns are useful to elucidate what elements are engaging in site competition. In Fig. 4(c) for the low energy facet in the quinary alloy, these patterns are especially clear in the elemental pair Cu/Fe. Cu enrichment reaches its maximum of ~39% at Plane -1 and Plane 1 and Fe its lowest value of approximately 10% at those same planes, accompanied by a small dip in Cu and increase in Fe concentrations in Plane 0. The Cu/Fe pair also swaps roles in Planes -3 and 3, which we note are outside what is traditionally identified as the grain boundary and have FCC structural order, where Fe becomes lightly enriched (~23%) and Cu mildly depleted (~18%). An analysis of the coordination numbers of atoms in each boundary plane indicate that Plane 0 atoms share the same coordination environment as atoms in the bulk, where both have 12 nearest-neighbor atoms. In contrast, atoms in Plane -1 and Plane 1 have a lower coordination number of 11. Since the peaks in Cu composition were observed at Plane -1 and Plane 1 instead of Plane 0, this provides evidence that Cu atoms on average appear to prefer grain boundary sites with more broken bonds (i.e., lower coordination), providing at least part of the driving force for segregation.

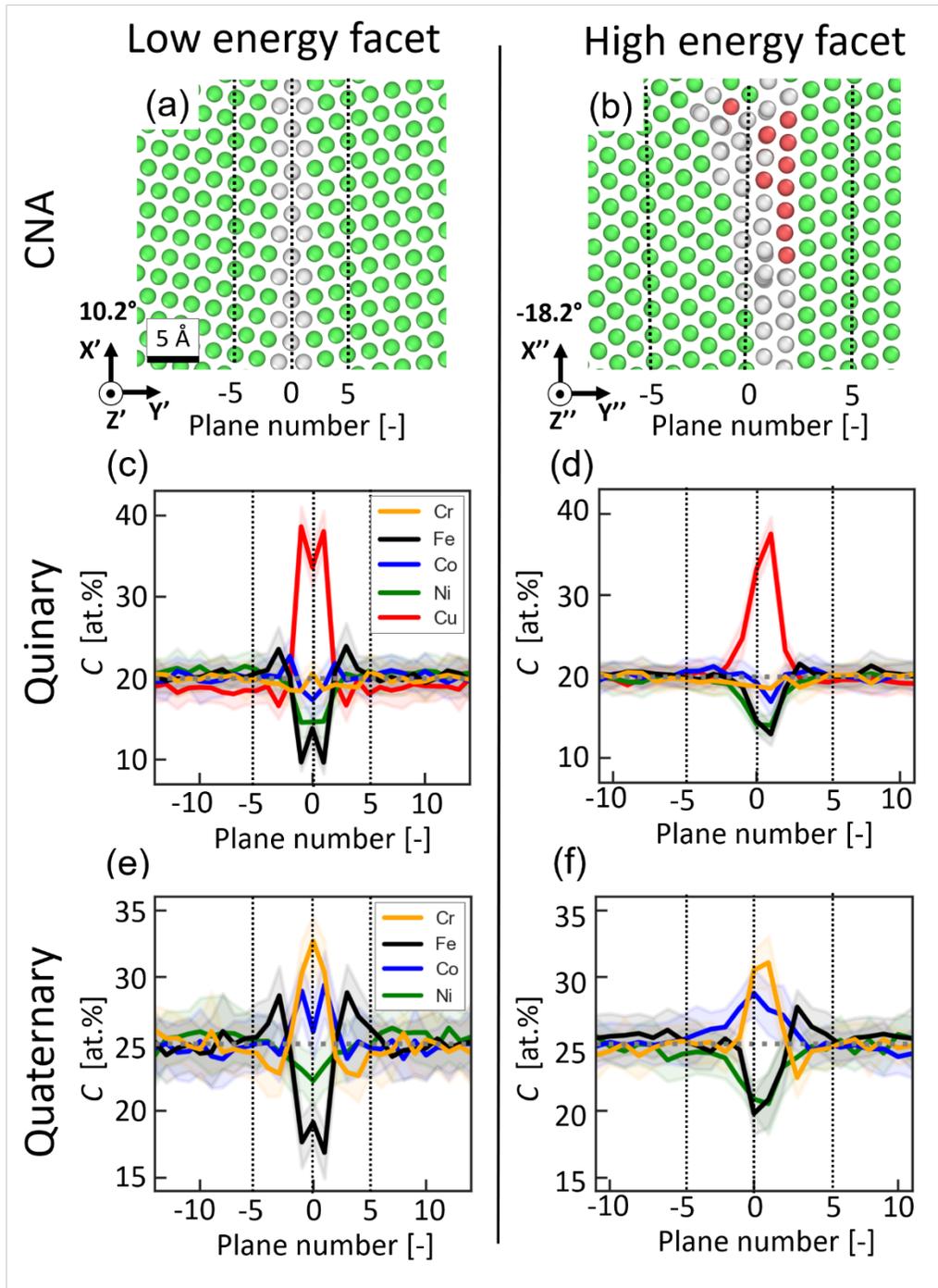

**FIG 4.** Plane-by-plane analysis of the two different types of facets at T = 1000 K. In the first row, an example of how the (a) low energy facets and (b) high energy facets are oriented for plane-by-plane analysis. Plane numbers -5 and 5 are shown on both for reference. (c, d) Concentration as a function of plane number for the quinary alloy. The dotted vertical lines are shown for reference and correspond to the plane numbers shown in (a) and (b). (e, f) The same analysis of the facets in the quaternary alloy.

The composition data for the high energy facets in the quinary alloy is shown in Fig. 4(d). As it is overall less ordered in its structure than the low energy facet, there are correspondingly fewer clear changes in composition. The right side of the facet, which has a high density of grain boundary defects, appears to resemble the aggregated Y-position data from Fig. 2(a). However, Plane 0 and Plane 1 are more ordered on average (evidenced by fewer in-plane positional fluctuations in Fig. 4(b)) and have only slightly lower maximum Cu enrichment values than seen in the low energy facet, while similar Cu/Fe coupling is also evident. Notably, the two terminal planes that comprise the high energy facet, namely (001) on the defect-heavy left and (111) on the more-ordered right, are natural contrasts with respect to intra-plane spacings of atoms (i.e., (001) is open while (111) is closest-packed). Thus, for this facet, the noted differences in segregation between highly ordered and more disordered planes is potentially influenced more heavily by differences in available free volume per plane.

The two facet types in the quaternary alloy are shown in Figs. 4(e) and (f). For the low energy facet, similar symmetric patterning of composition as seen in the quinary material is evident. Cr is the dominant grain boundary segregant in this case, reaching its maximum enrichment at Plane 0, followed by Co, which has its highest values of ~29 at.% at Planes -1 and 1. Fe depletion is also strongest at Planes -1 and 1 (concentration of ~16-17 at.%). Of note is clear Fe enrichment to ~28 at.% in Planes -3 and 3, which is beyond the grain boundary in an FCC-ordered region. The high energy facet (Fig. 4(f)) has significantly fewer symmetric trends in enrichment and depletion due to the facet structure, yet these behaviors are still clearly observed. Cr is once more the strongest segregant, followed by Co. Both Ni and Fe are depleted, reaching minimum concentration values of approximately 19 and 20 at.%, respectively. In both the low and high energy facets of the quaternary alloy, the pairing of Cr/Fe is inversely coupled to some extent, similar to the trends noted in the description of Figs. 4(c) and (d). This enrichment/depletion relationship is especially obvious in the regions outside of the boundary, specifically beginning at Planes -3 and 3 and extending through at least Planes -5 and 5, and even to approximately Plane 7 or 8 on the right side of the low energy facet in Fig. 4(e).

To examine whether spatial trends in CSRO exist, an analysis of the average $\alpha_I$ values (Equation 1) was conducted per plane on the low energy facets of each alloy, with the resulting curves shown in Fig. 5. Just as seen with the aggregated grain boundary CSRO data shown in Fig. 3, increases and decreases in $\alpha_I$ correspond strongly with the inverse trends in spatial concentration shown in Figs. 4(e) and (c) for the quinary and quaternary materials, respectively. Though this relationship is most strongly expressed in the dominant segregating elements for each material, the spatial CSRO data also highlights the noted regions of Fe enrichment in the planes adjacent to the low energy facets in the quaternary (Fig. 5(d)). Together, Fig. 3 and Fig. 5 confirm that little chemical ordering exists beyond that expected from changes in local elemental composition due to segregation, as might be expected from the interatomic potential's low mixing enthalpies.

On balance, two important trends emerge from the plane-by-plane analysis of elemental segregation. First, in the two symmetric low energy facets of Figs. 4(c) and (e), there are clear tendencies towards plane-based enrichment and depletion of certain elements within the grain boundary itself, such as that observed with the pair Cu/Fe in the quinary and Co/Fe in the quaternary. Such tendencies underscore findings that grain boundary character is an important factor in segregation in MPEAs [28,30]. Second, there is consistent observation of Fe enrichment in Planes -3 and 3 in all four examples (albeit to a subtle extent in Fig. 4(d)). Important to note is that these regions could be easily missed, even when using common analysis techniques for extracting the grain boundary region. For example, Planes -3 and 3 as well as the edges of facet junctions are in FCC-coordinated regions, so would not be identified as interfacial features by CNA.

A natural question that arises from these results is whether phenomena similar to the observed Fe enrichment of local FCC-ordered planes in the near-boundary region have been observed before. While we find no evidence of this type of behavior in MPEAs in grain boundaries in the literature, evidence of near-boundary segregation can be found in binary alloys and a few other chemically-complex material systems which support the trends observed in both facets. A study of Y segregation to a symmetric $\Sigma 3$ boundary in Y-stabilized zircona by Feng et al. [64] revealed a layered distribution of Y both in and around the boundary plane, a pattern attributed in part to the influence of grain boundary geometry. The symmetry of Y segregation seen in their results echoes that of Fe-segregation found in the low energy facet in this study. In a Ti-doped WC-Co cermet, Luo et al. [65] reported on highly unusual segregation phenomenon found at a mixed grain boundary. Specifically, segregation within the boundary itself was found to be dominated by W but was highly asymmetric with respect to composition and structure in the atomic layers neighboring the boundary plane, with Ti favoring one side and Co the other. These authors concluded that the nature of the terminating planes at the boundary was a major factor in determining the preferred segregating element and resulting structural changes. A similar result was observed by Kuo et al. [66] in a mixed character grain boundary in a Pt-3 at.% Ni alloy, where both oscillatory and monotonically decreasing Ni concentrations were observed on either side of a single interface. As the high energy facet studied here also has two different terminating planes (namely, $\{111\}_A/\{001\}_B$) and similarly asymmetric Fe enrichment, these results suggest that Fe segregation in FCC-ordered regions near the facets are also influenced by the terminating crystal planes' structure. Thus, the same tools that are used to analyze grain boundary segregation could also be applied to the near-boundary FCC regions to understand the source of Fe enrichment.

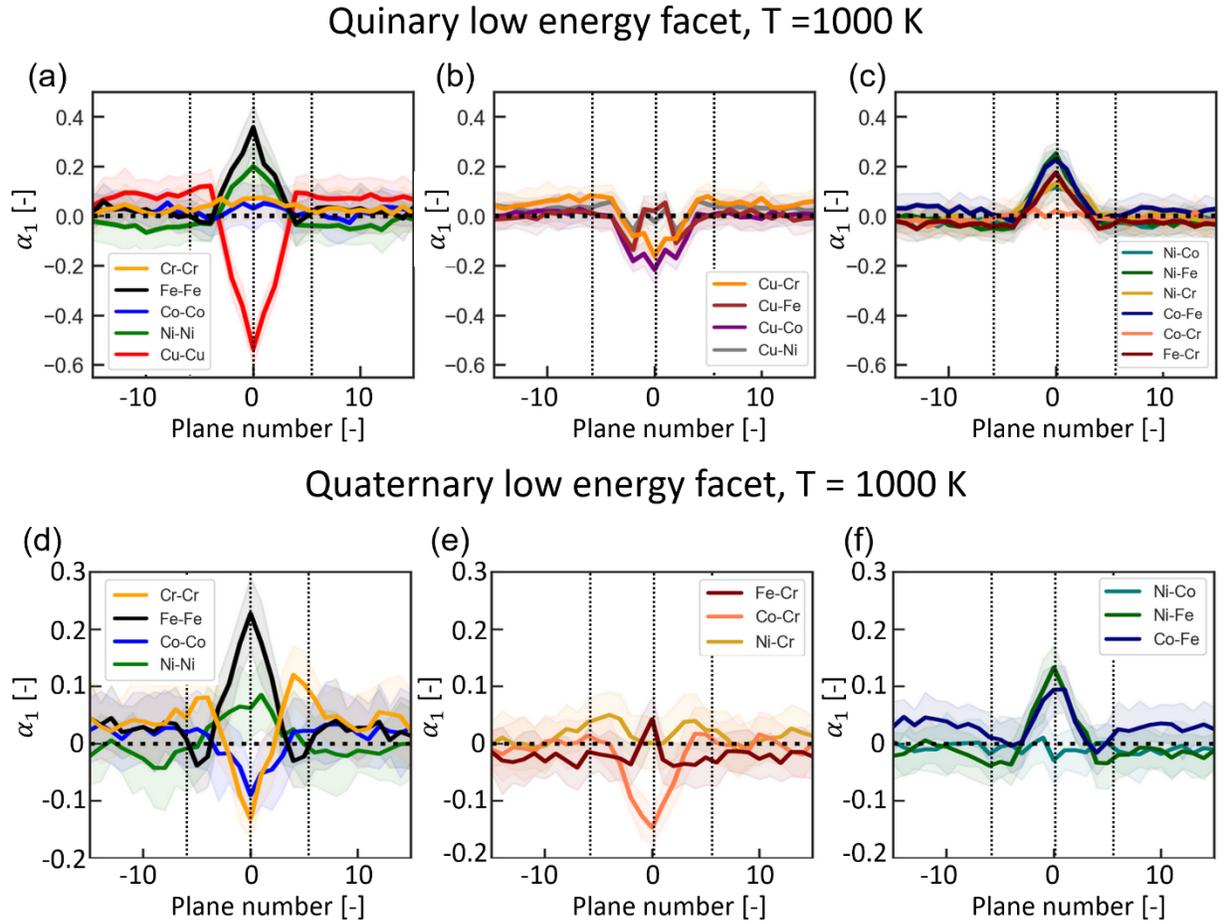

**FIG 5.** Analysis of trends in $\alpha_1$ as a function of plane number (similar to that conducted in Fig. 4) for the low energy facet only. The top row shows the data for the quinary alloy, with (a) same-element pairs, (b) Cu-based pairs, and (c) all other pairs. The bottom row shows the quaternary alloy data, with (d) same-element pairs, (e) Cr-based pairs, and (f) all other pairs. Dotted vertical lines indicate Planes 0, -5, and 5.

## 3.3 Grain boundary atomic volumes and segregation trends

The preceding two sections shed light on general and localized segregation trends. In this section, the property of atomic volume is used to explore connections between that segregation and grain boundary structure. Atomic volume (or the related parameter of atomic density) has been shown to be useful for understanding trends in boundary structure in simpler alloy systems. As an example, Fig. 6 shows the relationships between grain boundary structure and atomic volume $V_{atom}$ (units of Å$^3$) in the low energy facet in the quaternary material at T = 1000 K. Figure 6(a) shows a segment of these facets in detail. For the $\sum11$ <110> tilt misorientation, low energy facets consist of three plane layers. The boundary structure can also be seen as a chain of diamond-like features, two examples of which are outlined in the middle of Fig. 6(a). The vertices on the diamonds' long axis are linked to each other and together, form Plane 0 (dotted lines), while the outer two corners of each diamond touch along Planes -1 and 1 (dashed lines).

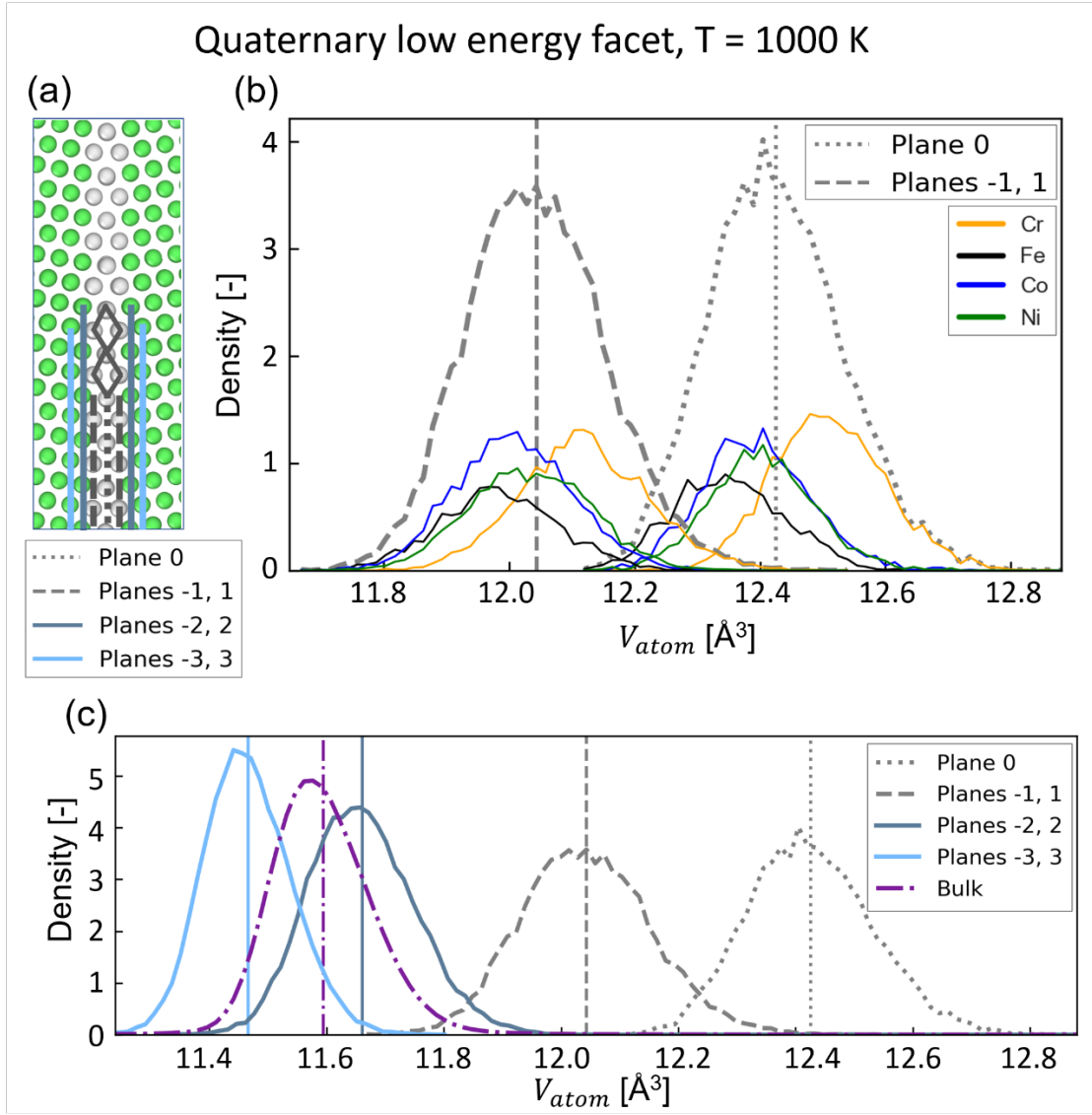

**FIG 6.** (a) The low energy facet type in the quaternary alloy at 1000 K, which can be characterized as a chain of diamond-shaped units (gray outline). Lines indicating the position of various planes are shown on the facet snapshot as well as in a legend beneath it. (b) The volume distributions for atoms in Plane 0 (dotted line) and Planes -1 and 1 (dashed lines), with both cumulative values (grey) and per-element quantities (various colors) presented. The mean of each distribution is shown with a vertical line. The distributions have been normalized so that their cumulative area is equal to one on these plots, since their absolute value is not critical (different planes have different numbers of atoms and therefore different densities) and the mean location along the X-axis is the critical feature. (c) The volume distributions, averaged across atom type, for Planes 0 through -3 and 3, as well as for the bulk crystal (purple). Unexpectedly, Planes -3 and 3 have mean atomic volumes lower than that of the bulk crystal.

In terms of atomic volume, the symmetry of the diamond shape means that atoms in Planes -1 and 1 are structurally identical to each other, and correspondingly distinct from Plane 0. Figure 6(b) shows histograms of atomic volume data (bin width: 0.015 $\text{Å}^3$), where two distinct gray curves (dotted and dashed) corresponding to each plane type (Plane 0, Plane -1/1) show the cumulative planar atomic volume data for that type and the colored curves show the per-element breakdown. The vertical lines for each plane type indicate each of their mean atomic volumes. To compare different planes to one another, the distributions are normalized so that the cumulative areas beneath each gray curve equal one. Within the boundary itself, Plane 0 has larger atomic volume than Planes -1 and 1. Similar planar atomic volume histograms are shown in Fig. 6(c) for other near-boundary planes for comparison, along with the average atomic volume distribution for bulk sites. Generally, the grain boundary sites (Planes -1, 0, and 1) have larger free volume than the bulk, as do Planes -2 and 2. An unexpected observation is that, at 11.46 $\text{Å}^3$, the mean volume of atoms in Planes -3 and 3 is noticeably lower than the bulk value (11.59 $\text{Å}^3$). Since these were in fact the planes where secondary enrichment of Fe was observed, the relationship between average planar volumes and average atomic volumes were investigated further.

Figures 7(a-d) show a plane-by-plane analysis of the average atomic volumes, $\bar{V}_{atom}$ in the quinary and quaternary alloys for the low and high energy facets (as before, on the left and

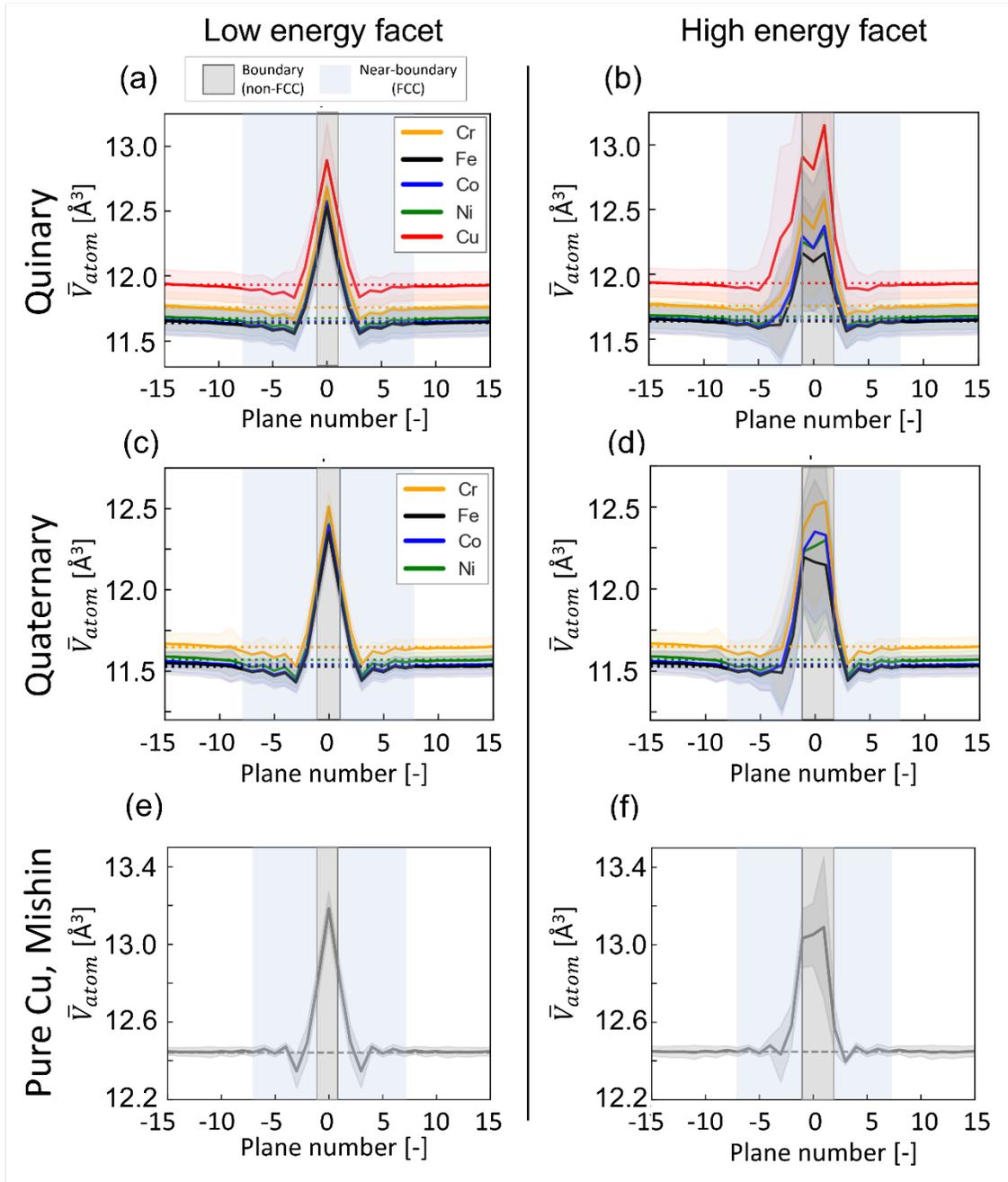

**FIG 7.** Atomic volume plotted as a function of plane number for both facet types at T = 1000 K. Dotted lines indicate the bulk average volume for each element. Gray shaded boxes outline the boundary (non-FCC) planes, and the blue shaded boxes define the near-boundary region (FCC), in which average atomic volumes are significantly different than bulk values. Data is shown for (a, b) the quinary alloy, (c, d) the quaternary alloy, and (e, f) pure Cu using a different interatomic potential from Mishin et al. [39].

right sides, respectively). The grain boundary planes (those identified with a majority of non-FCC atoms by CNA), are indicated using a shaded gray box. On average, the defects in the high-energy facet (Fig. 7(b) and 7(d)) were found to extend to approximately Plane -2. Dotted lines indicate the average atomic volume per element measured on atoms in FCC regions at least ±25 planes away from facets. Each element is indicated by different colors, and the dotted lines of the same colors indicate the average atomic volume in the bulk for that element, which were found to match the values found in a defect-free RSS cube for both MPEAs. As mentioned in Section 3.1 in reference to Equation 2, one way to study trends in MPEAs is to treat the per-element properties found in the bulk RSS as a reference to understand per-element changes at interfaces (in this case, average atomic volume). The numerical values for the bulk volumes per alloy are also included in Table 1.

Overall, it can be seen that the profiles of average atomic volumes for each element follow very similar contours within the interface and surrounding regions. For example, Figs. 7(a) and (c) show the data for the low energy facets in the quinary and quaternary alloys, respectively. All per-element volumes have their peak at Plane 0 in both alloys and have near-boundary regions with reduced atomic volumes, starting at Planes -3/3 and extending to Planes -8/8. Though within the interface there are subtle shifts in the magnitude of differences between element volume, the ordering remains very similar, following that seen in the bulk RSS material (Table 1). The same appears to be generally true in the high energy facets of Figs. 7(b) and (d), with some important caveats. The volume contours of the left and right near-boundary regions are significantly different. The right side, which is in general far more structured (see Figs. 1(c) and 4(b)), has its minimum volume value at Plane 3, which strongly resembles the pattern seen in the ordered low energy facets. In contrast, the near-boundary regions on the left side change more slowly, reaching their minimum values at approximately Plane -5 in both alloys. These differing profiles in the low and high energy facets are evidence of the strong influence of grain boundary character on per-element characteristics, even post-segregation. The elements themselves introduce a scaling factor, with relative magnitudes determined by local boundary geometry. It is also notable that the average atomic volumes in the MPEAs do not necessarily correspond to each element's average pure atomic volume shown in Table 1. For example, while Cr has the second-smallest average atomic volume in pure form, it has the largest volume after pure Cu when in MPEAs. It suggests that, in MPEAs, average pure atomic volumes can give some indication of segregation tendency when a species is relatively much smaller or larger (such as Cu), but other factors (such as surface energies) play an important role when pure atomic volumes are closer together in magnitude.

Though atomic volume (or relatedly, density) within the grain boundary plane has long been a topic of intense interest [67–70], exploration of the character of free volume in the bulk directly next to boundaries has seen less attention. In high resolution transmission electron microscopy, accurately determining grain boundary free volume expansion involves incorporating several atomic planes beyond what is considered the boundary itself [71]. Oscillations in plane distances have been observed in free surfaces in both pure metals and alloys [66,72]. However, to our knowledge, a distinct region of reduced atomic volume around grain boundaries as observed in these MPEAs has not been reported previously. We denote these as *near-boundary regions* and color them with light blue shading in Fig. 7. In order to understand whether this near-boundary region with altered atomic volume is present in pure materials as well, the same plane-by-plane analysis of the facets was conducted on simulations based on the pure Cu interatomic potential of Mishin et al. [39]. The pure Cu boundary was created using the same procedure outlined in the Methods section above. Figure 7(e) shows the per-plane atomic volume data for pure Cu atoms in the low energy facet type. The high symmetry of these facets is demonstrated again in the atomic volume data, which peaks at Plane 0 and drops rapidly until Planes -3/3 where, just like the MPEAs above, the atomic volume is reduced below bulk levels. The volume then oscillates around the mean over the following four or five planes. The same general pattern can be observed in the high energy facet data of Fig. 7(f) in both grains, despite this facet's less consistent structure (reflected in the larger error bands for the high energy facet). For both facets, volumes converge to the pure Cu bulk value by approximately Plane -7/7 on average. We do note that the width of the near-boundary region appears to be slightly smaller than that in the MPEAs, suggesting that this effect is material dependent. Very similar results were found when repeating this analysis in pure Cu using the MPEA potential. The fact that this near-boundary region of reduced and oscillating atomic volume exists in even a pure elemental system suggests that it is structural in nature, rather than a unique effect associated with the complex chemical composition of the MPEAs. In all of the boundaries and materials studied here, the presence of the grain boundary introduces a physically distinct zone of non-homogenous atomic volume that extends to the FCC-ordered planes neighboring the boundaries.

To better enable comparison between changes in atomic volume in the near-boundary region and local composition, magnified views of the atomic volume contours of the near-boundary regions of Figs. 7(a-d) are shown in Fig. 8. The solid lines show the per-element volume data for the fully-segregated structures, and the dashed lines show the same data for the RSS (chemically unrelaxed) structures for comparison. Because the high energy facets have higher symmetry on the right sides of their faces, those sides are also featured in Fig. 8. The shaded error bands and the dotted horizontal lines showing average atomic volumes per alloy from Fig. 7 have been removed for clarity. Beneath each volume plot are truncated versions of the composition versus plane plots of Fig. 4, with dotted black lines on Planes 3, 5, and 7 to guide the eye to common reference points. Plane 3 is the site of both maximum Fe enrichment and minimum average atomic volumes for three out of four cases, namely in the low energy facets of both alloys (Figs. 8(a) and (c)) and the high energy facet of the quaternary alloy (Fig. 8(d)). The enrichment extends beyond Plane 3 to roughly Plane 5 in the low energy facet of the quinary alloy and both facets of the quaternary alloy. In contrast, the near-boundary region of high energy facet of the quinary alloy in Fig. 8(b) has a slightly different trend compared to the other three near-boundary regions. That facet has only slight enrichment of Fe (~1.7 at.%) and no depletion of Cu, while the minimum atomic volume for Cu is found at Plane 5. This discrepancy is highlighted with the addition of a dark gray arrow and the thickening of the red curve between Planes 3 and 5 in Fig. 8(b). Comparison with the RSS curve (dashed line) shows that the structure's minimum volume was at Plane 3 after structural relaxation (annealing simulations using MD), but before chemical relaxation (simulations using MD/MC to find the final equilibrium state). Taken together, these trends indicate that the shift of the minimum to Plane 5 (solid line) is a result of chemical relaxation. Though further study of this exception is beyond the scope of this work, it may provide useful clues and a starting point for future investigations of near-boundary segregation.

Taken as a whole, the trends in the near-boundary regions described above provide an explanation for Fe enrichment in and beyond Plane 3, as well as the relative lack of enrichment in the high energy quinary facet of Fig. 8(b). Just as the free volume available at grain boundaries encourages segregation to the boundaries themselves, regions of reduced planar atomic volume nearby can also encourage segregation of elements. As the element with the vertical lines act as guidelines to compare trends in composition with those of RSS/segregated volume. Grey shading marks the grain boundary region, while blue shading marks the near-boundary region.

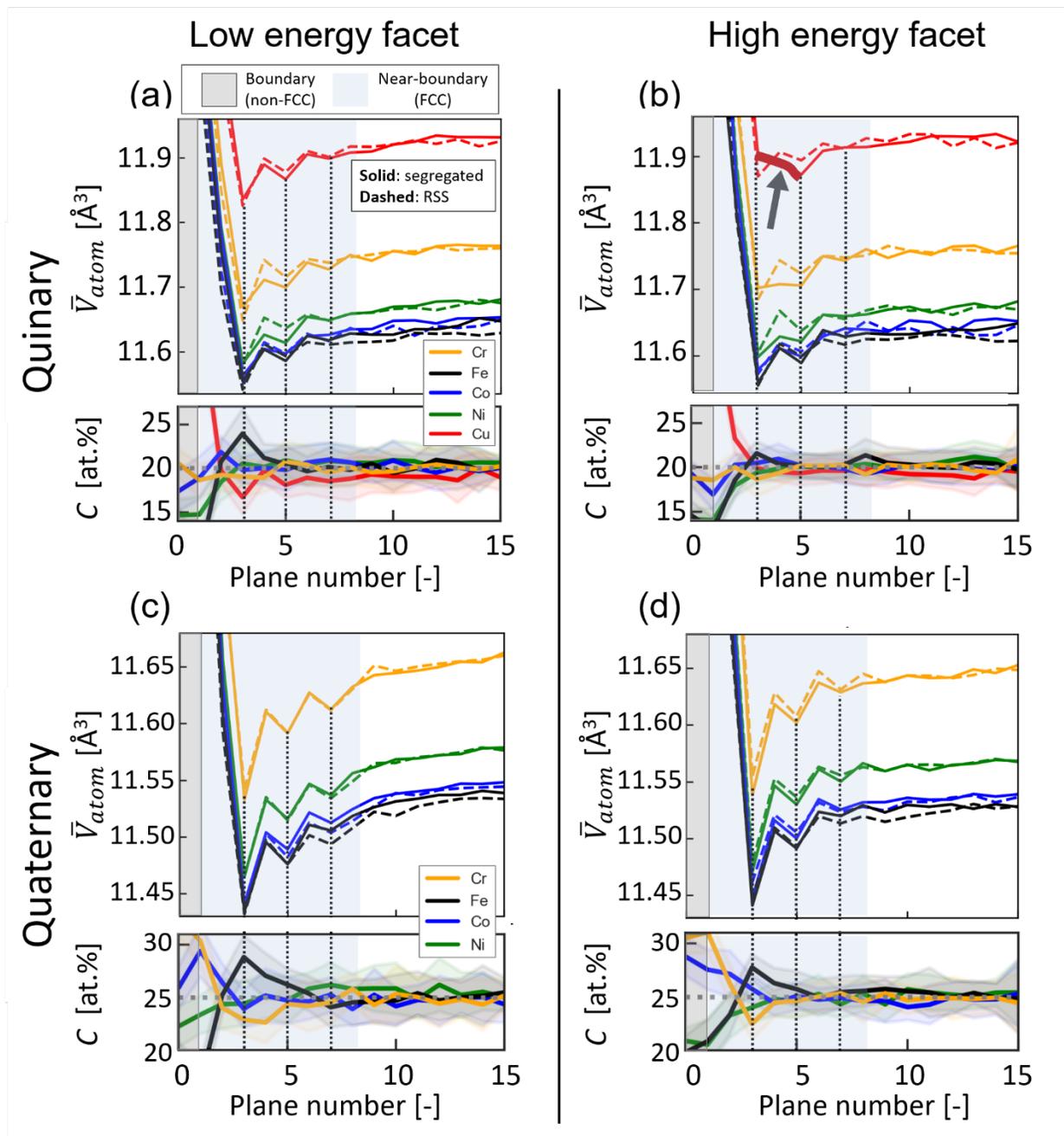

**FIG 8.** Magnified view of the average per-element atomic volumes for each plane in the near-boundary regions from Fig. 7 at T = 1000 K. Dashed colored lines show data for the random solid solution (RSS), and solid-colored lines show trends for the chemically relaxed, segregated state obtained by MD/MC simulation. Beneath each volume-plane plot is a truncated presentation to the composition-plane data from Fig. 4 for each material and facet. The dotted

smallest average atomic volume in both alloys, Fe is well-suited physically to fill the reduced planar volume sites in Plane 3. In contrast, the Cu atoms (with the largest atomic volume in all cases) are driven away from these sites, and their concentration depleted in these low planar volume regions. The same explanation applies to the quaternary alloy, where Fe segregates to and Cr depletes from sites with reduced planar volume in the near-boundary region.

As near-boundary segregation has been observed in both the symmetric low energy facet as well as the more disordered high energy one, we hypothesize that this effect is general and should be operative in other boundary types. Further atomistic study of grain boundaries in other MPEAs may even reveal near-boundary segregation to be ubiquitous. If so, this suggests that grain boundaries in MPEAs or other chemically complex materials in which site-competition effects are particularly pronounced might require a larger region to be analyzed than in chemically simpler materials. For example, it is common to model grain boundary widths as between 5 and 10 Å, while the combined region affected by the altered atomic volume (and segregation of different elements) is between approximately 11 and 17 Å here. Support for the idea of altered near-boundary regions in MPEAs specifically can be found in the models correlating segregation trends with a parameter that characterizes the degree of grain boundary disorder, utilized by Hu and Luo [33]. This disorder parameter can account for regions outside of what would traditionally be identified as the grain boundary when using methods such as Common Neighbor Analysis. Another interesting implication of near-boundary segregation is that it may provide an explanation for the origin of larger-scale phenomena, such as the strong near-boundary depletion observed by Zhao et al. [73]. It is possible that small but detectable changes in composition in the proximity of boundaries could act as chemical signals indicating potential sites of compositional transformation, such as the nanoclustering of Cr and Ni at grain boundaries observed in the Cantor alloy by Ming et al. [74].

## 4. Summary and Conclusions

Simulation and analysis of the segregation patterns at a faceted $\Sigma11$ <110> tilt boundary in two equiatomic MPEAs (CrFeCoNiCu and CrFeCoNi) allows the following conclusions to be made:

- Boundaries in both the quinary and quaternary alloys exhibit strong segregation. In the quinary alloy, dominant Cu segregation is driven by reductions in strain energy relative to the bulk and Cu's relatively low defect energies. The quaternary alloy has two strongly co-segregating elements, Cr and Co. Though their coupling complicates a precise analysis of the driving forces behind segregation, both elemental Cr and Co have defect energies that favor grain boundary sites over the bulk. Ni and Fe are strongly depleted at the grain boundaries in both MPEAs.

- Analysis of the CSRO shows that all atom pairs in the bulk have near-random chemical ordering after MD/MC simulation. Similar analysis of grain boundaries did not reveal any chemical ordering that could not be attributed to elemental composition changes due to segregation. Both results reflect the low enthalpies of mixing characteristic of this interatomic potential and highlight the importance of the other two driving forces described in this work, namely strain energy and boundary energy reduction.

- Significant segregation of Fe is observed in the FCC-ordered planes adjacent to both low and high energy facets, into what we term *near-boundary regions*. Atomic volume analysis reveals that this segregation is correlated with a structurally-distinct region of atomic volumes that are lower than the bulk. These near-boundary regions are present in pure Cu bicrystals as well, indicating that they are general features and not unique to interfaces in MPEAs. Fe segregation to the near-boundary region and its status as the atom with the smallest average atomic volume in both alloys suggests that other relatively small atoms may enrich low-density regions near interfaces as well. The fact that Fe does not have the smallest average atomic volume as a pure material suggests that the atomic size (measured through e.g. atomic volume, lattice parameter, or atomic radius) of the pure element alone may not necessarily predict its segregation behavior in the context of MPEAs.

The results from this study highlight an important gap in understanding segregation behavior, both in MPEAs as well as in more traditional alloy systems. Regions of crystalline material not traditionally identified as belonging to the grain boundaries but nonetheless structurally altered could impact microstructure and its evolution. Experimental techniques such as analytical transmission electron microscopy or atom probe tomography should be used in the near future to validate the presence of near-boundary segregation in MPEAs. The presence of near-interface segregation in cermets [65] and bulk metallic glasses [75] is an encouraging sign that such behavior is common in chemically-complex materials. Though near-boundary segregation zones may be especially obvious in MPEAs, they may have also been playing underappreciated roles in multi-component alloys, for example in ternary or even binary alloys beyond the dilute doping limit. In addition to providing fundamental insight into segregation behavior, such regions could be utilized in the tuning of material properties.

## Acknowledgements

This research was supported by the National Science Foundation Materials Research Science and Engineering Center program through the UC Irvine Center for Complex and Active Materials (DMR-2011967).